\documentclass[a4paper,11pt]{article}
\pdfoutput=1 

\usepackage{jcappub} 

\usepackage[T1]{fontenc} 

\title{\boldmath An Accurate Fitting Function For Scale-dependent Growth Rate in Hu-Sawicki $f(R)$ Gravity}

\author[a, 1]{Nareg Mirzatuny,\note{Corresponding author.}}
\author[a]{Elena Pierpaoli}

\affiliation[a]{Department of Physics and Astronomy, University Of Southern California,\\Los Angeles, California, USA}

\emailAdd{mirzatun@usc.edu}
\emailAdd{pierpaol@usc.edu}

\abstract{We present a simple fitting formula for the scale-dependent growth rate of Hu-Sawicki model in $f(R)$ modified gravity. We compare the accuracy of the fitting function against numerical results and report achieving a sub-percent maximum error for all different studied cases. The validity of the fitting function is tested against a wide range of scales, $10^{-4} \le k \le 5 \ [\mathrm{Mpc}^{-1}]$ , for various redshift values in the range of $z \in [0.0,\ 3.0]$ and different $|f_R^0|$ values. This formula is useful in producing predictions for cosmological probes that are sensitive to growth rate such as redshift space distortions, galaxy-galaxy correlation function, peculiar velocity statistics and pairwise velocities of galaxy clusters.}

\begin{document}

\maketitle

\flushbottom

\section{Introduction}

\label{sec:intro}

Since its unexpected discovery more than 20 years ago \cite{Riess:1998cb, Perlmutter:1998np} the accelerated expansion of the Universe has remained in the spotlight as one of the major unsolved problems of modern physics. Cosmological measurements of distant type Ia supernovae were the original indicators of the existence of this phenomenon. Subsequently, these discoveries were followed by observations in cosmic microwave background anisotropies and large scale structure surveys, all supporting the aforementioned discovery \cite{Aghanim:2018eyx}. The standard paradigm of modern cosmology, $\Lambda \mathrm{CDM}$ model, explains the late time accelerating expansion of the Universe by virtue of an additional constant, $\Lambda$, added to the Einstein field equations. However, the discrepancy between the observed and the theoretically expected value for the cosmological constant leads to different phenomenological and philosophical problems \cite{Carroll:2000fy, Weinberg:1988cp}. Hence, several theoretical approaches and new frameworks have been proposed to solve this problem \cite{Joyce:2016vqv}. The majority of the developed solutions belongs to one of the two different paradigms: 1 - Dark Energy models, where exotic new forms of energy are introduced to source the late time accelerated expansion of the Universe. The cosmological constant can be classified as a specific member of this group. 2 - Modified Gravity models, that are based on the assumption of the inadequacy of General Relativity (GR) to explain the dynamics of the gravitational field on certain cosmological scales. Here, the theory of gravity changes and typically a new degree of of freedom is introduced in the theory that plays the role of the cosmological constant naturally \cite{Dvali:2000hr}.

The distant supernovae observations and the background expansion history alone are not sufficient to distinguish between different Dark Energy and Modified Gravity models \cite{Sahni:2006pa}. On the other hand the growth of cosmological perturbations has been shown to have the potential to discriminate between the two models \cite{Piazza:2013pua, Mirzatuny:2013nqa}. Upcoming surveys like Euclid, DESI, WFIRST, LSST \cite{Amendola:2012ys, Aghamousa:2016zmz, Ivezic:2008fe} will map huge volumes of the Universe and the structure within, opening the door for detailed analysis of the structure formation over the last three quarters of the history of the Universe \cite{Huterer:2013xky} which in turn will give the researchers the opportunity to better discriminate between alternative models to cosmological constant scenario.

Motivated by these arguments, we present a fitting formula for the scale-dependent growth rate of $f(R)$ gravity. The fitting function is capable of reproducing the numerical results in redshifts and scales that will be measured accurately by the upcoming and future surveys. Previous attempts to model the growth rate of modified gravity with analytical or semi-analytical approaches have been able to reproduce acceptable results \cite{Tsujikawa:2009ku}. Here we present a new fitting function that is capable of reproducing the numerical results up to very small scales, namely $10^{-4} \le k \le 5 \ [\mathrm{Mpc}^{-1}]$ for the redshift range of $z \in [0.0, 3.0]$ with a sub-percent accuracy. In addition, the explicit dependence of the fitting formula on $|f_R^0|$ provides a simple but accurate way of using the growth rate in future cosmological analysis.
 
 This paper is organized as follows. In section 2 we briefly review the $f(R)$ gravity and the specific model that we have used in this study. Section 3 is a short review of structure formation for $\Lambda \mathrm{CDM}$ and $f(R)$ gravity with the emphasise being on the growth rate. In section 4 we present the fitting formula for the growth rate of matter density perturbations and scrutinize it against different cosmological scenarios. Section 5 is dedicated to potential applications of the fitting function in large-scale structure cosmology. Lastly, Section 6 is our final remarks and conclusions. Throughout the paper we have used $H_0 = 70\ [\mathrm{km\ s^{-1} Mpc^{-1}}] $ while assuming a flat Friedmann-Lemaitre-Robertson-Walker (FLRW) background.




\section{$f(R)$ Gravity Review}

A potential model for explaining the late time accelerated expansion of the Universe is modified gravity. Among different proposed models, $f(R)$ modified gravity is of particular interest. By employing the chameleon mechanism \cite{Mota:2003tc, Khoury:2003rn} these theories modify General Relativity on cosmological scales while at the same time passing the local tests of gravity. In the $f(R)$ model of gravity the Einstein-Hilbert action is replaced by
\begin{equation}
    S = \int d^4 x\sqrt{-g}\bigg[\frac{f(R)}{16\pi G} + \mathcal{L}^{(m)}\bigg],
\end{equation}
where $G$ is the Newton's gravitational constant and $\mathcal{L}^{(m)}$ is the matter Lagrangian \cite{Nojiri:2003ft, Starobinsky:2007hu}. $f(R)$ is a function of the Ricci scalar, $R$ and the specific functional form is determined by a combination of theoretical and phenomenological considerations. In the special case of $f(R) = R - 2\Lambda$ one recovers the Einstein-Hilbert action plus cosmological constant and consequently $\Lambda \mathrm{CDM}$ cosmology. Modified Einstein field equations have the following form in $f(R)$ gravity:
\begin{equation}
    FR_{\mu \nu} - \frac{1}{2}fg_{\mu \nu} + (g_{\mu \nu}\Box - \nabla_{\mu}\nabla_{\nu})F = 8\pi G T_{\mu \nu},
\end{equation}
where 
\begin{equation}
    F(R) \equiv \frac{\partial f(R)}{\partial R}.
\end{equation}
$F$ is a new degree of freedom of the theory that acts similar to a scalar field, dubbed $\textit{scalaron}$, with a mass given by
\begin{equation}
    M^2 = \frac{1}{3}\bigg(\frac{F}{F_{,R}}\label{mass}\bigg),
\end{equation}
where $F_{,R} = \partial^2 f(R)/\partial R^2$. The scalaron field propagates on scales that are smaller than its associated comoving Compton wavelength,
\begin{equation}
    \lambda_c = a^{-1}\sqrt{3\frac{\partial F(R)}{\partial R}}.
\end{equation}
Modified gravity models in general are able to reproduce and mimic the redshift-distance measurements of supernove and are indistinguishable from $\Lambda \mathrm{CDM}$ and Dark Energy models in this regard \cite{Appleby:2007vb}. Therefore, "background"-level observations alone are not sufficient to test or falsify these models. Despite that, in linear perturbation regime the equations governing the relationship between matter overdensities and the gravitational potential are also modified, leading to a different growth characteristic and growth history. Hence modified gravity have unique and falsifiable predictions in this context and can be tested against future observations. In particular, $f(R)$ models modify the growth of density perturbations and structure formation via introduction of a scale-dependant growth which is in contrast to $\Lambda \mathrm{CDM}$ and the scale-independent growth therein \cite{Tsujikawa:2009ku}.

At linear perturbation level, in the comoving gauge the modified Einstein equations lead to the following perturbation equations in Fourier space for the evolution of matter overdensities:

\begin{equation}
    \ddot{\delta}_m + \bigg(2H + \frac{\Dot{F}}{2F}\bigg)\dot{\delta}_m - \frac{\rho_m}{2F}\delta_m =  \frac{1}{2F}\bigg[\Big(-6H^2 + \frac{k^2}{a^2}\Big)\delta F + 3H\dot{\delta F} + 3\ddot{\delta F} \bigg], \label{perturb1}
\end{equation}
\begin{equation}
    \ddot{\delta F} + 3H\dot{\delta F}+\bigg(\frac{k^2}{a^2} + \frac{F}{3F_{,R}} - \frac{R}{3}\bigg)\delta F = \frac{1}{3}\rho_m \delta_m + \dot{F}\dot{\delta}_m\ . \label{perturb2}
\end{equation}
%
Here $k$ is the comoving wavenumber, $a=(1+z)^{-1}$ is the scale factor normalized to unity for the current time, $\rho_m$ is the matter density, $\delta_m (a) = \delta \rho_m/\rho_m$ is the density contrast,  $H = \dot{a}/a$ is the Hubble parameter and "$\ ^{.}\ $" represents derivative with respect to the cosmic time. 
Since we are interested in the evolution of density perturbations in $f(R)$ gravity we combine eq. $eq. ~ \eqref{perturb1}$ and $eq. ~ \eqref{perturb2}$ to obtain the modified equation of evolution for the matter density contrast. For $f(R)$ models that are cosmologically viable, F has small variation ($|\dot{F}| \ll HF$), consequently in both equations $\dot{F}$ can be neglected. The oscillating mode is also negligible compared to modes induced by matter perturbations. Finally since we are interested in modes well within the Hubble radius we have $k^2/a^2 \gg H^2$. All the above assumptions leads to the following equation,
\begin{figure}[]
	\centering 
	\includegraphics[width=0.75\textwidth]{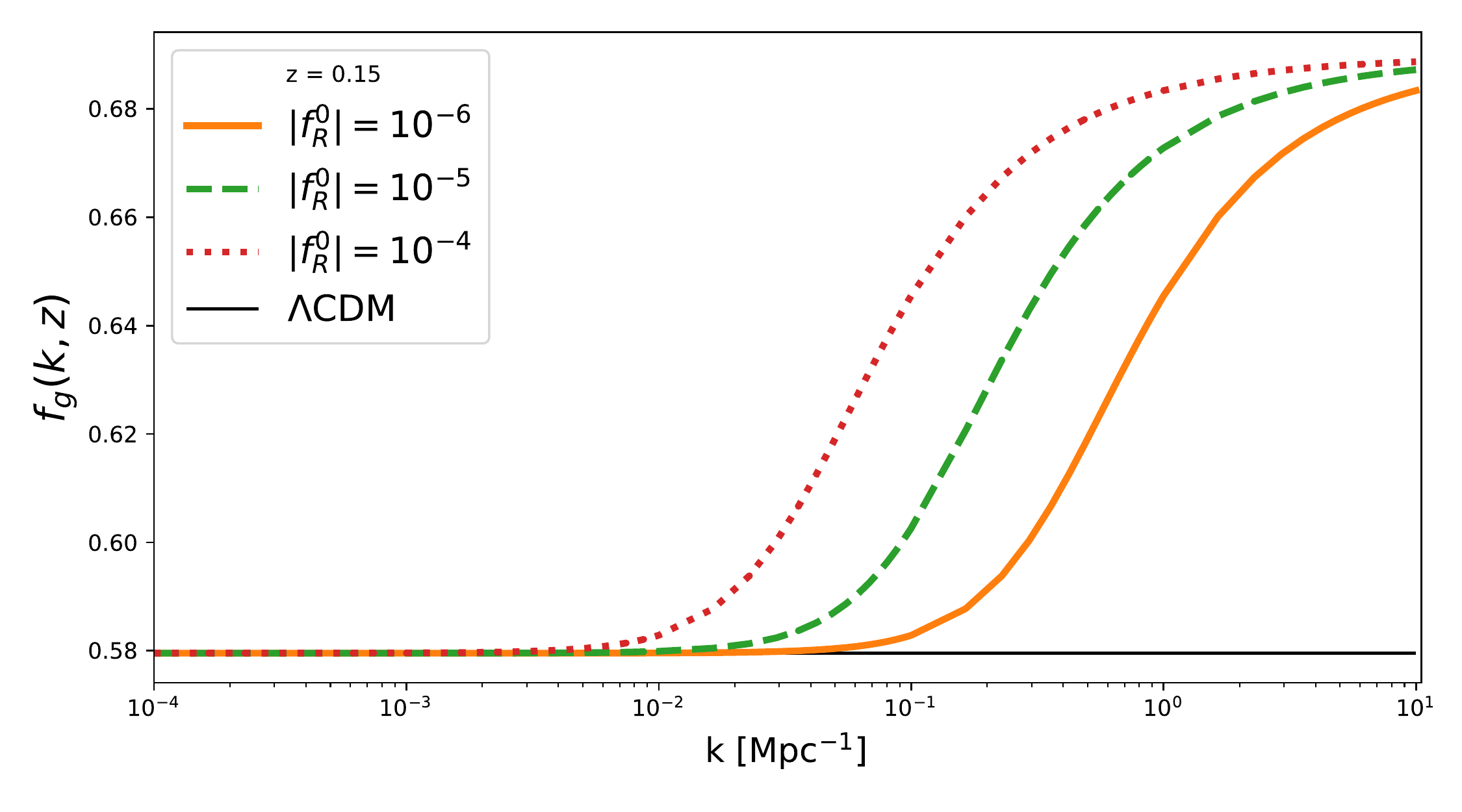}
	\caption{Comparison of growth rate for $\Lambda \mathrm{CDM}$ and $f(R)$ gravity with different values for $|f_R^0|$ at a given redshift. The the scale-dependent features in growth rate of $f(R)$  gravity become more prominent on the small scales.}
\end{figure}

\begin{equation}
    \ddot{\delta}_m + 2H\Dot{\delta}_m - 4\pi G_{\mathrm{eff}}\rho_m \delta_m \simeq 0\label{eq:2.6},
\end{equation}
where $G_{\mathrm{eff}}$ is the "effective" gravitational constant, defined as
\begin{equation}
    G_{\mathrm{eff}}(k, z) = \frac{G}{F}\bigg[1 + \frac{1}{3}\Big(\frac{k^2}{a^2 M^2/F + k^2}\Big)\bigg].
\end{equation}
Since $RF_{,R} \ll 1$, $eq. ~ \eqref{mass}$ can be rewritten as $M^2 = F/3F_{,R}$.
It is worth noting that $G_{\mathrm{eff}}$ has replaced $G$, which would instead appear in standard $\Lambda \mathrm{CDM}$ cosmology. Since $G_{\mathrm{eff}}$ is a scale-dependent function, the structure formation in modified gravity includes scale dependent features that are unique and can in principle be used to distinguish $f(R)$ models from standard cosmology \cite{Narikawa:2009ux}.
%

So far we have not assumed any specific functional form for $f(R)$. A viable model should be able to satisfy a wide range of cosmological and local constraints imposed by observational and theoretical considerations. In this work we use the Hu-Sawicki $f(R)$ model (with $n=1$) that has been previously shown to pass the local solar system tests \cite{Hu:2007nk}. The functional form for this model is as follows,

\begin{equation}
    f(R) = R - 2\Lambda  \frac{R}{R + \mu^2},\label{hu}
\end{equation}
where $\Lambda$ and $\mu$ are free parameters. With the assumption of small deviations from GR ($R \gg \mu^2$), $eq. ~ \eqref{hu}$ can be simplified as:
\begin{equation}
    f(R) = R - 2\Lambda - f_R^0\frac{\bar{R}_0^2}{R},\label{eq:2.11}
\end{equation}
where $|f_R^0| = 2\Lambda \mu^2 /\Bar{R}_0^2$. Here, $R = 6(2H^2 +\dot{H})$ and $\bar{R}_0 = \bar{R}(z=0)$ where overbar indicates a background quantity.
Using eq. $\eqref{eq:2.11}$, the comoving Compton wavelength for this model is
\begin{equation}
    \lambda_C = (1+z) \sqrt{6|f_R^0|\bigg( \frac{\bar{R}_0^2}{R^3}\bigg)}.
\end{equation}
It is conventional to use $|f_R^0|$ to quantify the deviation of $f(R)$ models from General Relativity. A smaller value of $|f_R^0|$ is an indication of weaker deviation. Latest cosmological studies indicate that $|f_R^0| \leq 10^{-5}$ \cite{Cataneo:2014kaa, Liu:2016xes}. Regardless, we have assumed a wide range of potential values for $|f_R^0| = \{10^{-6}, 10^{-5}, 10^{-4}\}$ in this work. The inclusion of $|f_R^0| =10^{-4}$ in our analysis is for demonstration purposes only.
\begin{figure}[]
	\centering 
	\includegraphics[width=1\textwidth]{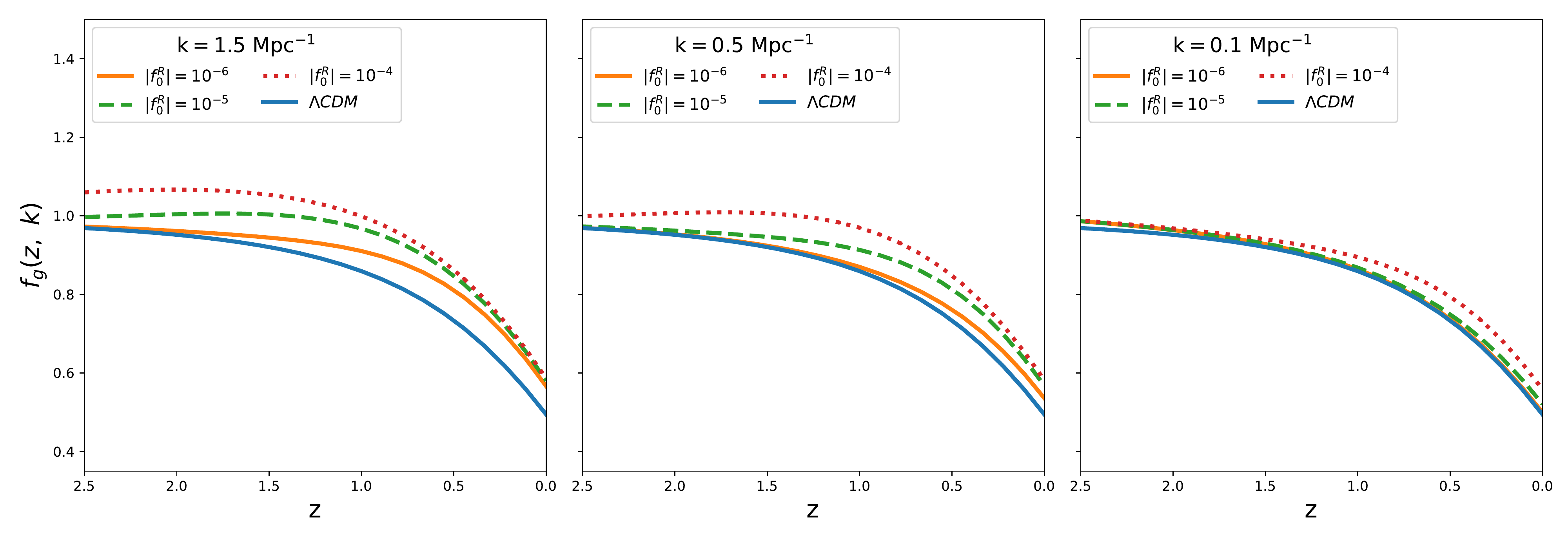}
	\caption{Growth rate vs. redshift for $\Lambda \mathrm{CDM}$ and $f(R)$ gravity. Three different scales for modified gravity is considered, $k \in \{0.1, 0.5, 1.5\}\ [\mathrm{Mpc}^{-1}]$. As been shown in the previous plot, the deviation from $\Lambda \mathrm{CDM}$ is more prominent for the larger $k$-modes (smaller scales).}
\end{figure}
\section{Growth Rate of Matter Density Perturbations}

The growth rate of matter density perturbations in $\Lambda \mathrm{CDM}$ cosmology is a function of scale factor alone \cite{Peebles:1984ge, Lahav:1991wc} and is defined as,
\begin{equation}
    f_g(a) = \frac{d \ln{\delta_m (a)}}{d \ln{a}}.
\end{equation}
\begin{figure}[!t]
\centering
        \includegraphics[width = 1.0\textwidth]{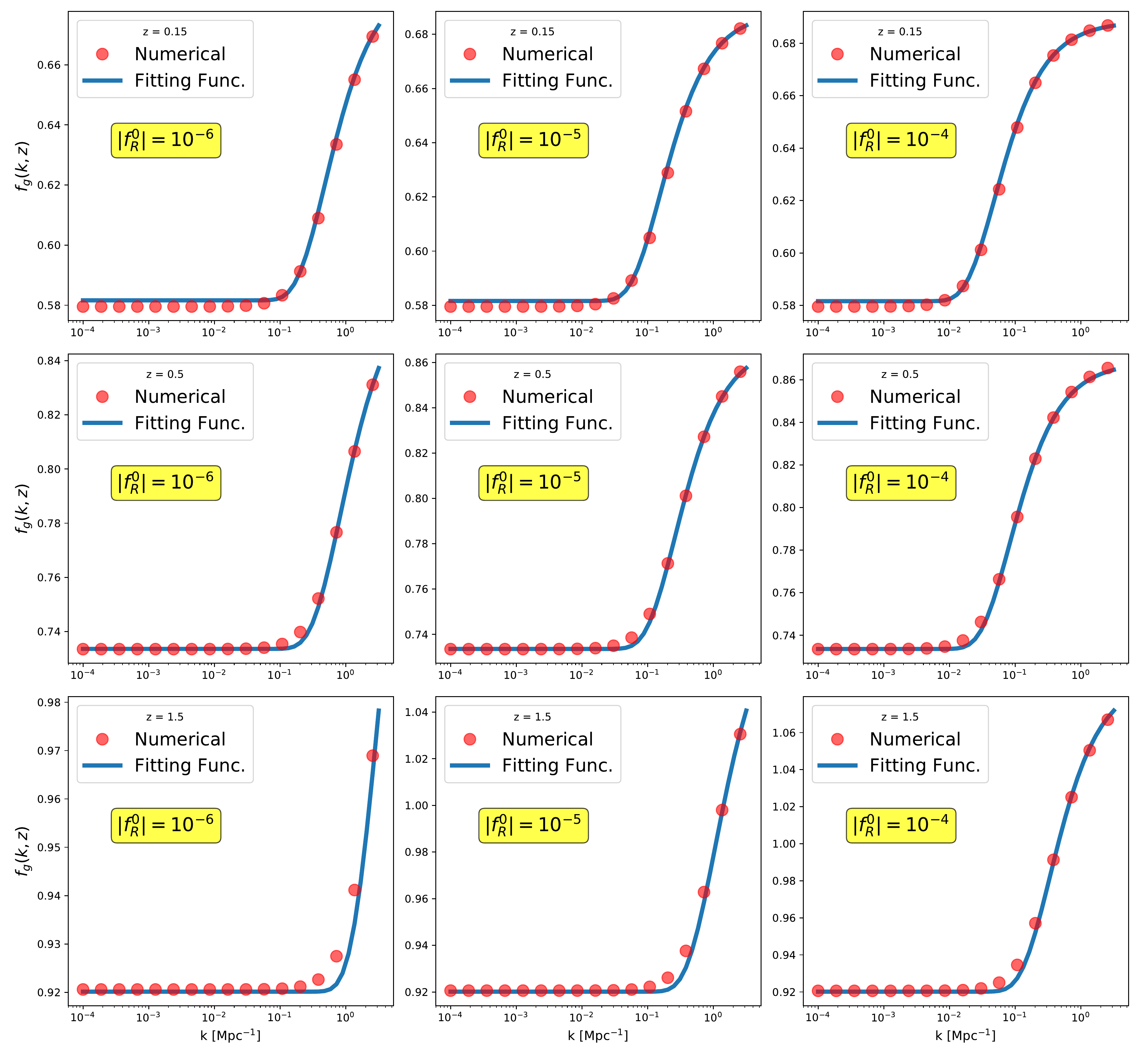}
        \begin{center}
        \caption{Comparison of numerical results with fitting formula for three different choices of redshift, z = $\{0.01, 0.5, 1.5\}$. We plot the exact numerical solution of equation (3.4) (red dots) along with our fitting function, equation 4.1. The fitting function accurately regenerates the numerical results on all scales with a sub-percent maximum error. The fitting function works equally well for different $|f_R^0|$ values.}
        \end{center}
\end{figure}
It has been shown that for $\Lambda \mathrm{CDM}$ cosmology, to a high level of accuracy, $f_g(a)$  can be approximated by
\begin{equation}
    f_g(a) = \Omega_m^{\gamma}(a).
\end{equation}
where $\gamma \simeq 0.55$ is the growth index and $\Omega_m (a) = \Omega_{m, 0} a^{-3} E^{-2}(a)$ is the matter density for flat FLRW cosmology \cite{Yin:2019rgm} with $ E(z) = \frac{H(z)}{H_0} = \sqrt{\Omega_{m,0} (1+z)^3 + \Omega_{\Lambda}}$.
In the case of $f(R)$ gravity, due to the scale-dependent nature of $G_{\mathrm{eff}}(k, z)$, the growth rate is expected to have scale-dependent features as well and can be written as,
\begin{equation}
    f_g(k, a) = \frac{d \ln{\delta_m (k, a)}}{d \ln{a}}.\label{fgfR}
\end{equation}
Since our aim is to study the growth rate of matter perturbations in $f(R)$ gravity, using $eq. ~ \eqref{fgfR}$ we can rewrite eq. $eq. ~ \eqref{eq:2.6}$ in terms of $f_g$ and in a form that is numerically more tractable,
\begin{equation}
f_g^{\prime}(k, z) - \frac{f_g^2 (k, z)}{1 + z} + \Big(\frac{E^{\prime}(z)}{E(z)} - \frac{2}{1 + z}\Big)f_g(k, z)  +  \frac{3}{2}\Omega_{m}(z)(1+z)\frac{G_{\mathrm{eff}}(k, z)}{G} = 0,\label{eq:3.4}    
\end{equation}
where we have changed the differentiation variable from cosmic time to redshift, $^{\prime} \equiv \partial /\partial z$.
By solving $eq. ~ \eqref{eq:3.4}$ numerically, in Fig. 1 we show the scale-dependant behavior of growth rate in $f(R)$ gravity. The figure shows that deviations from $\Lambda \mathrm{CDM}$ become larger on small scales. In addition, a larger value of $|f_R^0|$ produces larger deviations. In Fig. 2 we plot the growth rate against redshift while having the scale fixed. As evident, the deviation from GR is a combination of redshift  and scale.
\section{Fitting Function For Growth Rate}
\begin{figure}[!t]
\centering
        \includegraphics[width = 0.8\textwidth]{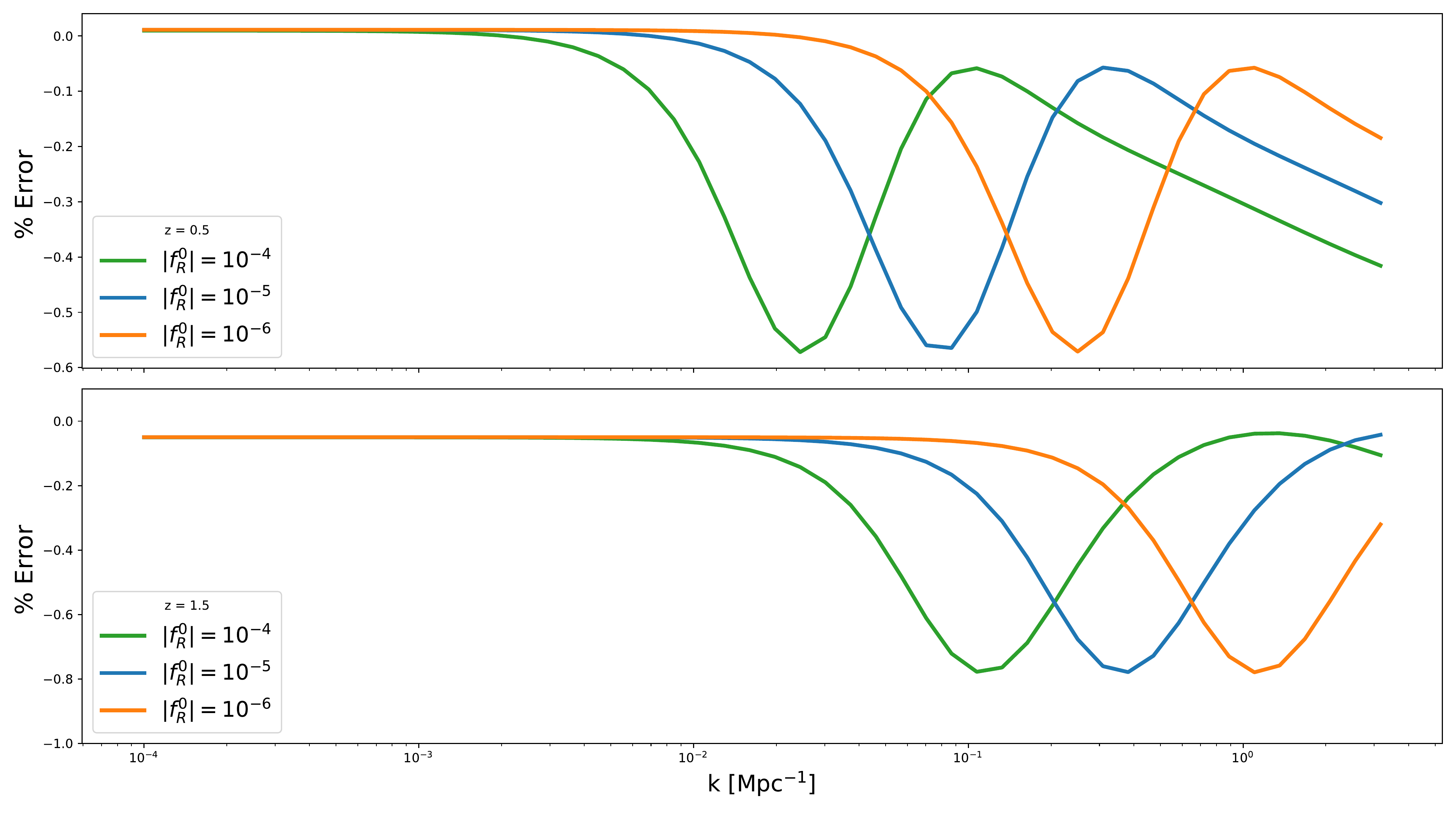}
        \begin{center}
        \caption{$\Delta f_g(\%)$ vs. $k$ for different $|f_R^0|$ values at $z = \{ 0.5, 1.5 \} $. As evident in the plot the difference between the numerical results and the fitting formula is below 1\% for all the included scales.}
        \end{center}
\end{figure}
 Analytical and/or semi-analytical approaches to model the scale-dependent growth rate for modified gravity has been proposed before \cite{Resco:2017jky, BuenoSanchez:2010wd, Narikawa:2009ux, Denissenya:2017uuc}. We propose a simple fitting function that is capable of recreating the numerical results with high accuracy on a wide range of scales for various redshifts while having an easy to use functional form.
We have calculated $f_g$ over a range of values for $|f_R^0|$ across different redshifts. Our intention is to find a fitting function that properly reproduce the numerical results not only on larger scales but also on small ones. Consequently, we present the following fitting formula 
\begin{equation}
    \large
    f_g(k, z;\ |f_R^0|) = \Omega_m ^{\gamma}(z)\bigg(1+\frac{e^{\frac{-\alpha(z)}{\mathcal{K}}}}{5.5}\bigg)\label{ff},
\end{equation}
where we have defined the new variable $\mathcal{K}$ as
\begin{equation}
    \mathcal{K} \equiv k\lambda_C.
\end{equation}
As before, $k$ is the comoving wave number and $\lambda_C$ is the comoving range of the scalar field force, defined in the previous section. $\alpha(z)$ is a redshift dependent functions defined as
\begin{equation}
    \begin{aligned}
            \alpha(z) & = 1.35 + 0.52z -0.15z^2.\\
    \end{aligned}
\end{equation}
The truncation of $\alpha(z)$ at second order in terms of redshift is based on the observation that a linear truncation is not sufficient enough to produce the required accuracy (below 1$\%$ error), while at the same time, inclusion of a cubic term has no meaningful contribution to the results.
We evaluated the accuracy of the fitting formula for different values of redshift, scale, $\Omega_m$ and $|f_R^0|$. Namely, we have considered cases where $z \in [0.0, 3.0]$, $|f_R^0| \in \{10^{-6}, 10^{-5}, 10^{-4}\}$ and $10^{-4} \le k \le 5\ \mathrm{[Mpc^{-1}}]$ with $\Omega_m \in [0.28, 0.335]$. We report that the fitting function is able to match the numerical results by a sub-percent accuracy for all the studied cases. Fig. 3 is a demonstration of the quality of the fitting function when compared against a large collection of numerical results. It is worth mentioning that the explicit dependence of the fitting formula on $|f_R^0|$ makes the computation of derivatives with respect to $|f_R^0|$ and consequently potential use in Fisher Matrix analysis straightforward.
We assess accuracy of the fitting formula against numerical results by
 \begin{equation}
     \Delta f_g (\%) \equiv (\frac{f_g^{\mathrm{fit.\ func.}} - f_g^{\mathrm{numerical}}}{f_g^{\mathrm{numerical}}}) \times 100.
 \end{equation}
 In Fig. 4, we plot the relative error of the fitting formula against numerical results. As evident in Fig. 4, the fitting formula has a sub-percent accuracy over a very wide range of scales for $z \in [0.0,\ 3.0]$ for different $|f_R^0|$ values. Here $\Omega_{m,0} = 0.2815$ has been used.
 
 \section{Cosmological Probes and Applications}
In this section we briefly discuss cosmological probes that are sensitive to the growth rate directly or indirectly where the fitting formula can be used to perform calculations for cosmological analysis.
\subsection{Redshift Space Distortion}
\begin{figure}[!]
        \includegraphics[width = 1.0\textwidth]{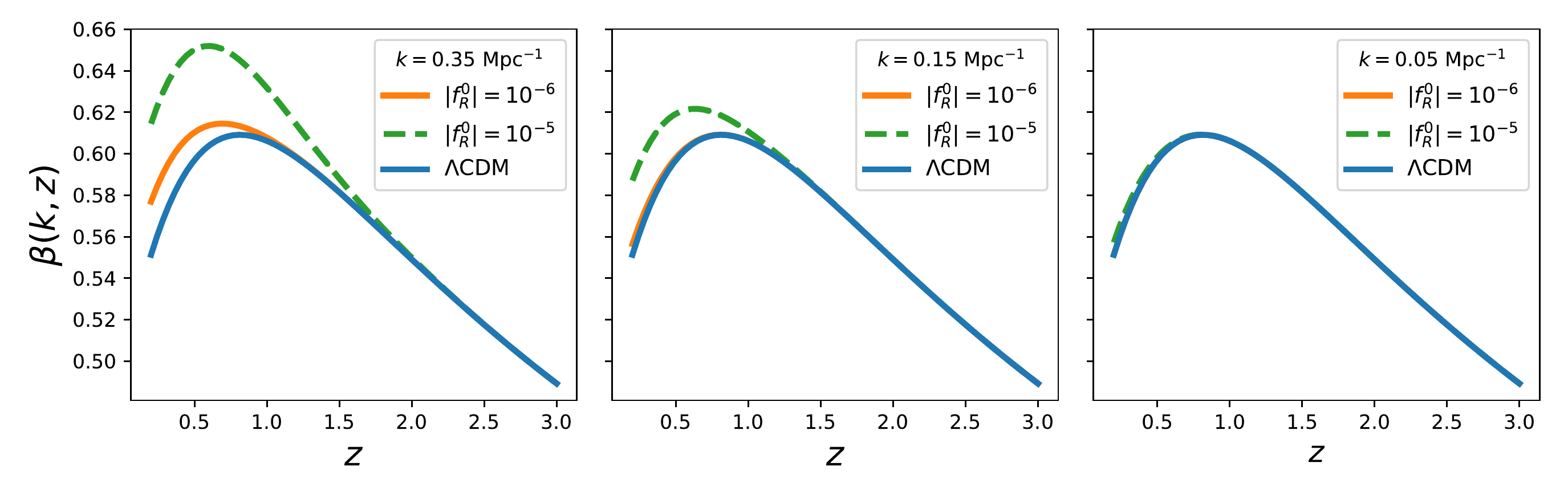}
        \caption{Redshift space distortion parameter, $\beta\ vs \ z.$ Here we show the scale dependent feature of $\beta(k, z)$ for $f(R)$ gravity. Two different values for deviation from $\Lambda \mathrm{CDM}$ have been considered, $|f_R^0| = \{10^{-5}, 10^{-6}\}$. As evident in the above plots, the larger deviations from $\Lambda \mathrm{CDM}$ is expected at larger $k$ modes.}
\end{figure}
Clustering patterns of galaxies in redshift space is affected by their peculiar velocities. On large scales this phenomenon is called the Kaiser effect \cite{Kaiser:1987qv}: Infalling of galaxies into regions with high-density such as galaxy clusters. On small scales, the Fingers-of-God effect is present: The clustering pattern of galaxies is elongated along the line of sight in the redshift space, caused by random peculiar velocities of galaxies that are bound in virialised structures. In linear perturbation theory, the relation between the redshift space galaxy power spectrum and the power spectrum in real space is as follows \cite{Jackson:2008yv},
\begin{equation}
    P_s (k, \mu) = (1 + \beta \mu^2 )^2 P_r (k).
\end{equation}
Here $\mu = \frac{\bold{k}\cdot \bold{r}}{kr}$ is the cosine of the angle between the line of sight and direction of infall. The distortion parameter, $\beta$, is defined as
\begin{equation}
    \beta(k, z) = \frac{f_g(k, z)}{b(z)},
\end{equation}
where $b(z) = \delta_g/\delta_m$ is the linear galaxy bias. In contrast with $\Lambda \mathrm{CDM}$, because of the scale dependent nature of growth in $f(R)$ gravity the distortion parameter is scale dependent too. In Fig. 5 we plot the distortion parameter against redshift for different $k$ modes for various $|f_R^0|$ values. We have assumed a simple and only redshift-dependent bias, $b(z) = \sqrt{1+z}.$
\subsection{Velocity Correlation Function}
In linear perturbation theory Fourier components of peculiar velocity and density contrast field are related as follows \cite{peebles},
\begin{equation}
    \bold{v}(\bold{k}) = -iH_0f_g \frac{\bold{\hat{k}}}{k}\delta_m (\bold{k})\label{eq::5.1}.
\end{equation}
\begin{figure}[!]
        \includegraphics[width = 1.0\textwidth]{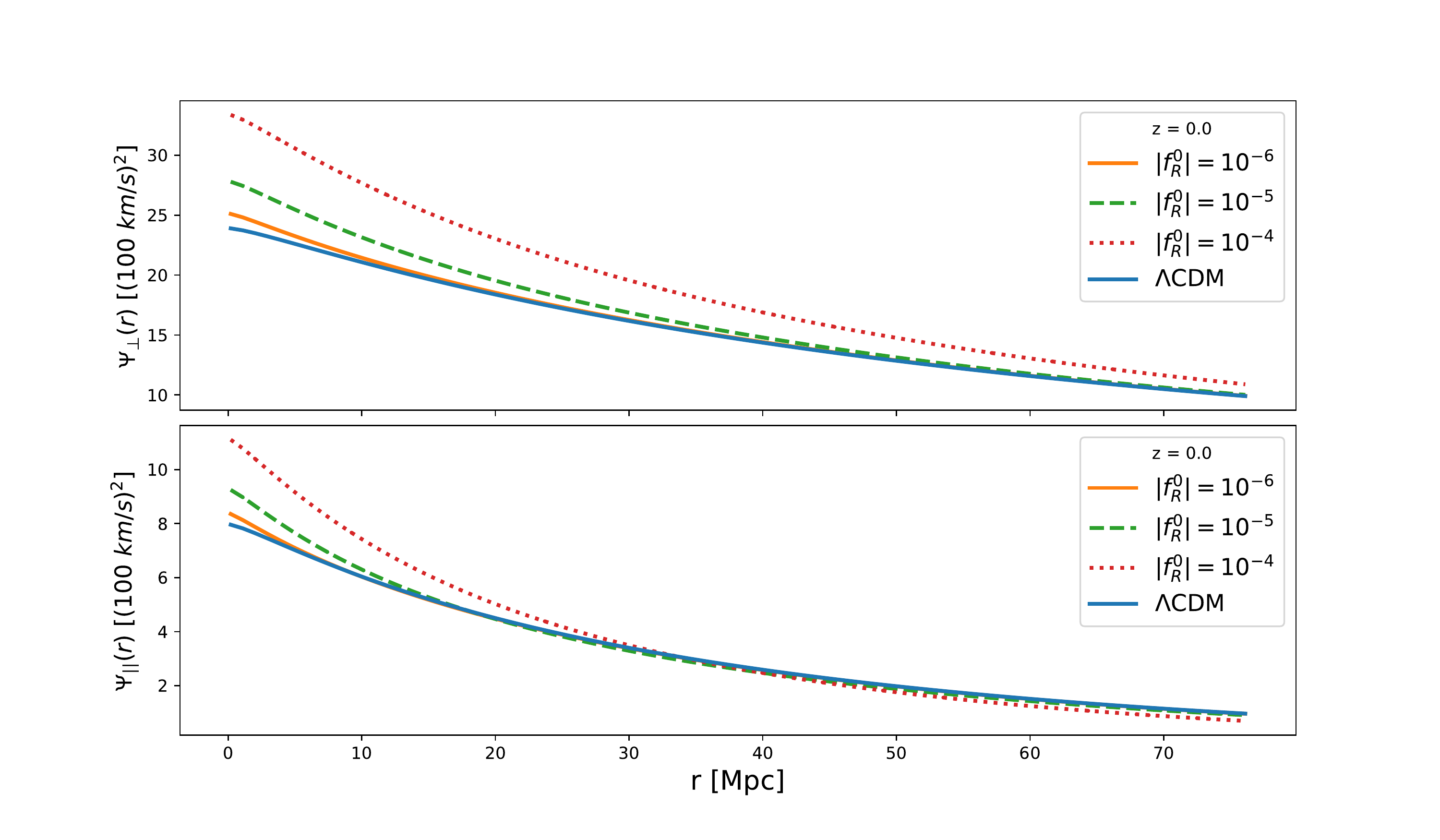}
        \caption{Top: In this plot we present a comparison between $\Psi_{\bot}(r)$ for $\Lambda \mathrm{CDM}$ and $f(R)$ gravity for different values of $|f_R^0|$. We have included $|f_R^0| = 10^{-4}$ for comparison purposes only. As expected the deviation becomes more dominant on smaller scales. Bottom: Same as above with $\Psi_{\bot}(r)$ being replaced by $\Psi_{||}(r).$}
\end{figure}
In the context of $f(R)$ modified gravity since the energy-momentum conservation equations are unchanged, eq. \eqref{eq::5.1} is not altered and can be used. In order to study the correlation of peculiar velocity field the velocity correlation tensor is defined as,
\begin{equation}
    \Psi_{ij}(r)\equiv \langle v_i(\bold{x})v_j (\bold{x}+\bold{r})\rangle,
\end{equation}
where $\bold{r}$ is the separation vector between two points located at $r_i$ and $r_j$. For a statistically homogeneous and isotropic velocity field the correlation tensor is a linear combination of the parallel, $\Psi_{||}$ and transverse (to the separation vector), $\Psi_{\bot}$ components \cite{gorski1:1988, gorski2:1989},

\begin{equation}
    \Psi_{ij}(r) = \Psi_{\bot}(r)\delta_{ij} + \big[\Psi_{||}(r) - \Psi_{\bot}(r)\big]\hat{r}_i \hat{r}_j
\end{equation}
where $\delta_{ij}$ is the Kronecker delta. It has been shown that given the matter power spectrum, $P(k)$, $\Psi_{ ||, \bot}$ can be calculated directly as

\begin{equation}
    \Psi_{\bot}(r) = \frac{H_0^2}{2\pi^2}\int dk f_g^2(k,z) P(k)\frac{j_1 (kr)}{kr},
\end{equation}
and
\begin{equation}
    {\Psi}_{||}(r) = \frac{H_0^2}{2\pi^2}\int dk f_g^2(k, z) P(k)\bigg[j_0 (kr) - 2\frac{j_1 (kr)}{kr} \bigg],
\end{equation}
where $j_0 (x) = \frac{sin(x)}{x} \hspace{0.5mm}$ and $j_1 (x) = \frac{sin(x)}{x^2}-\frac{cos(x)}{x}$ are the spherical Bessel functions.
We like to emphasize the fact that in standard $\Lambda \mathrm{CDM}$ case the scale-independent growth rate, $f_g(z)$ is not part of the integration. Since $\Psi_{||}$ and $\Psi_{\bot}$ are functions of power spectrum and growth rate, in principle one can study the parallel and transverse components of the velocity correlation tensor to constrain modified gravity. In Fig. 5 we have plotted $\Psi_{\bot}(r)$ and $\Psi_{||}(r)$ for $\Lambda \mathrm{CDM}$ and $f(R)$. As expected the $f(R)$ curves start deviating from GR on smaller scales. We have used the publicly available MGCAMB code \cite{Hojjati:2011ix, Zhao:2008bn} to calculate the $f(R)$ power spectrum for different $|f_R^0|$ values.

\subsection{Pairwise Velocity of Galaxy Clusters}

The recent discovery of pairwise kinematic Sunyaev-Zel'dovich effect \cite{Hand:2012ui} is a new opportunity to use the second order CMB physics to study gravity. pkSZ can act as a proxy to study pairwise velocity of galaxy clusters and it has been shown to have great potential to constrain gravity and dark energy models on cosmic scales \cite{Mueller:2014nsa}. We are planning to address this problem in a more detailed fashion in an upcoming paper \cite{NM2019}.
\subsection{Galaxy-Galaxy Angular Correlation}

By ignoring the very small angular multipoles, namely for $\ell \ge 10$ and using Limber approximation \cite{limber} the galaxy-galaxy angular auto-correlation can be simplified as,
\begin{equation}
    C^{g_ig_j}(\ell) = \int dz \frac{H(z)}{\chi^2 (z)}b_i W_{g_i}(z)b_j W_{g_j}(z)[D_{m}^2 (k, z)P(k)]_{k = \frac{\ell+1/2}{\chi(z)}}\label{gg}.
\end{equation}
The growth function, $D_{m}(k, z)$, is related to the growth rate and can be expressed as
\begin{equation}
    \int d\ln D_m(k,z)=-\int dz \frac{f_g(k,z)}{1+z},
\end{equation}
where $f_g(k, z)$, the fitting formula can be used to perform the calculation. In equation \eqref{gg}, $\chi (z)$ is the comoving distance to redshift $z$, $b_i$ is the galaxy bias for the redshift bin $i$ and $W_{g_i}(z)$ is the galaxy redshift distribution in the same bin (normalized to 1). In addition to the galaxy-galaxy angular auto-correlation, the fitting formula can be used for calculation of angular cross-correlation between Cosmic Microwave Background temperature and galaxy distribution (Integrated Sachs-Wolfe cross-correlated with galaxies). Here, on top of using the fitting formula for obtaining $D_m(k, z)$, the integrated Sachs-Wolfe effect needs to be modeled in modified gravity context \cite{Schmidt:2008hc}.

\section{Conclusions}
Future high precision large-scale structure surveys will map the distribution and evolution of matter densities in the Universe with unprecedented detail and will provide opportunities to study gravity and alternative models of gravity on cosmological scales.
In this work we reviewed the scale-dependant growth of large-scale structure in $f(R)$ gravity under different assumptions about the underlying model and for various redshifts. Motivated by the potential of the future experiments to constrain modified gravity, we introduced a fitting formula for scale-dependent growth rate of matter perturbations that is in sub-percent agreement with numerical results over a wide range of scale, redshift and has explicit dependence on $|f_R^0|$, the parameter indicating deviation from $\Lambda \mathrm{CDM}$ cosmology. We discussed number of cosmological probes that directly or indirectly can be used to study the growth rate. Namely, the fitting formula is able to regenerate the numerical results with high accuracy for $10^{-4}\le k \le 5\ [\mathrm{Mpc}^{-1}]$, $0 \le z \le 3$ for $|f_R^0| = \{10^{-6}, 10^{-5}, 10^{-4}\}$. Using the fitting formula, we calculated the peculiar velocity statistics for $f(R)$ gravity and compared the results to standard cosmology. Depending on the scale that is considered and the value for $|f_R^0|$ the $f(R)$ predictions can deviate from standard cosmology significantly. Lastly, we reviewed the redshift space distortion phenomena in the context of $f(R)$ gravity while using the fitting formula for obtaining the results.

\acknowledgments
We like to thank Shant Baghram and  Siavash Yasini for helpful discussions. NM and EP are supported by NASA grant 80NSSC18K0403. NM was partially supported by the WiSE major support for faculty, awarded to EP. The authors acknowledge using the \texttt{SciPy} library \citep{Scipy} for preparation of the results presented in this work. 

\bibliographystyle{plain}	

\bibliography{cosmobib.bib}

\end{document}